\documentclass{article}
\usepackage{graphicx}

\def\alt{\mathrel{\rlap{\lower3.5pt\hbox{$\mathchar"218$}}\raise 2pt
                \hbox{$\mathchar"13C$}}}
\def\agt{\mathrel{\rlap{\lower3.5pt\hbox{$\mathchar"218$}}\raise 2pt
                \hbox{$\mathchar"13E$}}}
\newcommand\Rey{\mbox{\it Re}}  

\begin{document}
\setcounter{page}{1}

\title{Pressure determinations for incompressible fluids and magnetofluids}
\author{Brian T. Kress and David C. Montgomery\\ \small{Department of Physics and Astronomy, Dartmouth College,}\\ \footnotesize{Hanover, NH 03755-3528, USA}}
\date{}
\maketitle

\begin{center}
To appear in Journal of Plasma Physics\\
\end{center}
\vspace{1pc}

\begin{abstract}

Certain unresolved ambiguities surround pressure determinations for 
incompressible flows, both Navier-Stokes and magnetohydrodynamic. For 
uniform-density fluids with standard Newtonian viscous terms, taking the divergence of the equation of motion leaves a Poisson equation for the 
pressure to be solved. But Poisson equations require boundary conditions.  
For the case of rectangular periodic boundary conditions, pressures determined in this way are unambiguous. But in the presence of  ``no-slip" rigid walls, the equation of motion can be used to infer both Dirichlet and Neumann boundary 
conditions on the pressure $P$, and thus amounts to an over-determination. This has occasionally been recognized as a problem, and numerical treatments of wall-bounded shear flows usually have built in some relatively {\it ad hoc} dynamical recipe for dealing 
with it, often one which appears to ``work" satisfactorily.  Here we consider a 
class of solenoidal velocity fields which vanish at no-slip walls, have all 
spatial derivatives, but are simple enough that explicit analytical solutions for $P$ can be given. Satisfying the two boundary conditions separately gives two pressures, a ``Neumann pressure" and a ``Dirichlet pressure" which differ non-trivially at the initial instant, even before any dynamics are implemented. We compare the two pressures, and find that in particular, they lead to different volume forces near the walls. This suggests a reconsideration of no-slip boundary conditions, in which the vanishing of the tangential velocity at a no-slip wall is replaced by a local wall-friction term in the equation of motion.

\end{abstract}

\section{Introduction}

It has long been the case that pressure determinations for incompressible 
flows, both Navier-Stokes and magnetohydrodynamic (MHD), are known to be highly non-local. Taking 
the divergence of the equation of motion
\begin{equation}
\frac{\partial {\bf v}}{\partial t} + {\bf v} \cdot \nabla{\bf v} = 
\frac{{\bf j}{\times}{\bf B}}{\rho c} - \nabla P + \nu \nabla^2{\bf v},
\label{eq: eqmo}
\end{equation}  
and using $\nabla \cdot {\bf v} = 0$ leaves us with a Poisson equation for 
the pressure $P$, which is said to function as an equation of state:
\begin{equation}
\nabla^2 P = -\nabla\cdot({\bf v}\cdot\nabla{\bf v}-\frac{{\bf j} 
{\times}{\bf B}}{\rho \it c}) 
\label{eq: pe}
\end{equation} 
Here, ${\bf v}={\bf v}(x,t)$ is the fluid velocity field as a function 
of position and time, ${\bf B}$ is the magnetic field, ${\bf j}=c\nabla{\times}{\bf B}/4\pi$ is the electric current density, $c$ is 
the speed of light, $\nu$ is the kinematic viscosity, assumed spatially 
uniform and constant, and $P$ is the pressure normalized to $\rho$ the mass density, also spatially uniform. (\ref{eq: eqmo}) and (\ref{eq: pe}) are written for MHD. Their Navier-Stokes equivalents can be obtained simply by dropping the 
terms containing ${\bf B}$ and ${\bf j}$.

If we are to solve (\ref{eq: pe}) for $P$, boundary conditions are 
required. In the immediate neighborhood of a stationary ``no-slip" boundary, 
both the terms on the left of (\ref{eq: eqmo}) vanish and we are left 
with the following equation for $\nabla P$ as a boundary condition:
\begin{equation}
\nabla P = \nu\nabla^2{\bf v}+\frac{{\bf j}{\times}{\bf B}}{\rho c}
\label{eq: bcs}
\end{equation}	
								
We now focus on the Navier-Stokes case, where the magnetic terms disappear 
from (\ref{eq: bcs}), for simplicity. All the complications of MHD are illustrated by this simpler case. It is  apparent that (\ref{eq: bcs}) must apply to all components of $\nabla P$, and that while the normal component of $\nabla P$ is enough to determine $P$ through Neumann boundary conditions, the tangential components of (\ref{eq: bcs}) at the wall equally well determine $P$ through Dirichlet boundary conditions. This is a problem which some inventive procedures have been proposed to resolve, usually by some degree of ``pre-processing" or various dynamical recipes which seem to lead to approximately no-slip velocity fields after a few time steps (e.g.  \cite{Gresho91}, \cite{Roache82} and \cite{canuto88} ). It is not our purpose to review or critique these recipes, but rather to focus on a 
set of velocity fields, related to Chandrasekhar-Reid functions 
\cite{Chandrasekhar61}, for which (\ref{eq: pe}) is explicitly soluble 
at a level where the Neumann or Dirichlet conditions can be exactly 
implemented. In~\S\,\ref{sec:pdetermin}, we explore the difference between 
the two pressures so arrived at. Then in~\S\,\ref{sec:discussion}, we 
propose a replacement for the long-standing practice of demanding that all 
components of a solenoidal ${\bf v}$ vanish at 
material walls, in favor of a replacement by a wall friction term for which 
the above mathematical difficulty is no longer present. Of course, similar statements and options will apply to all comparable incompressible MHD problems.

\section{Pressure determinations}\label{sec:pdetermin}

Restricting attention at present to the Navier-Stokes case, we consider two-dimen- sional, solenoidal, velocity fields obtained from the following stream function:
\begin{equation}
\psi(x,y) = C_{k\lambda}\cos{(kx)}[\cos{(\lambda y)} + A_{k\lambda}\cosh{(ky)}]
\label{eq: sf}
\end{equation}
The hyperbolic cosine term in (\ref{eq: sf}) contributes a potential 
flow velocity component to ${\bf v}$ which makes it possible to demand that $\bf v$ obey two boundary conditions: the vanishing of both components at rigid walls \cite{Chandrasekhar61}. The function in (\ref{eq: sf}) is even in $x$ 
and $y$, but can obviously be converted into an odd or mixed one by the 
appropriate trigonometric substitutions.

The velocity field ${\bf v} = \nabla\psi{\times}\hat{{\bf e}}_z$ has 
only $x$ and $y$ components and is periodic in $x$, with an arbitrary 
wavenumber $k$. $C_{k\lambda}$ is a normalizing constant, and the constants 
$\lambda$ and $A_{k\lambda}$ can systematically be found numerically to any desired accuracy so that both 
components of ${\bf v}$ vanish at symmetrically placed no-slip walls at $y = a$ 
and $y = -a$.  In fact, for given $k$, an infinite sequence of such pairs 
of $\lambda$ and $A_{k\lambda}$ can be determined straightforwardly. Thus any 
such ${\bf v}$, or superposition thereof, is not only solenoidal, but has both 
components zero at $y = \pm  a$, and all spatial 
derivatives exist. Moreover, the ``source" term, or $\nabla\cdot({\bf 
v}\cdot\nabla{\bf v})$, from the right hand side of (\ref{eq: pe}), is 
of a relatively simple nature for such a ${\bf v}$, since every term in it can be 
written as a product of exponentials of $kx$, $\lambda y$ and $ky$.  It is 
straightforward to find an inhomogeneous solution for $P$, which then is 
the same for all boundary conditions for a given ${\bf v}$ of the form stated. To 
this inhomogeneous part of $P$ must be added a solution of Laplace's 
equation.  This can be chosen so that the total $P$ may satisfy either the 
normal component of (\ref{eq: bcs}) at the walls, or the tangential 
component of it, but not both.  The determination involves only simple but 
tedious algebra. 

\begin{figure}[!ht]
  \begin{center}
  \includegraphics[width=\textwidth,height=3.5in]{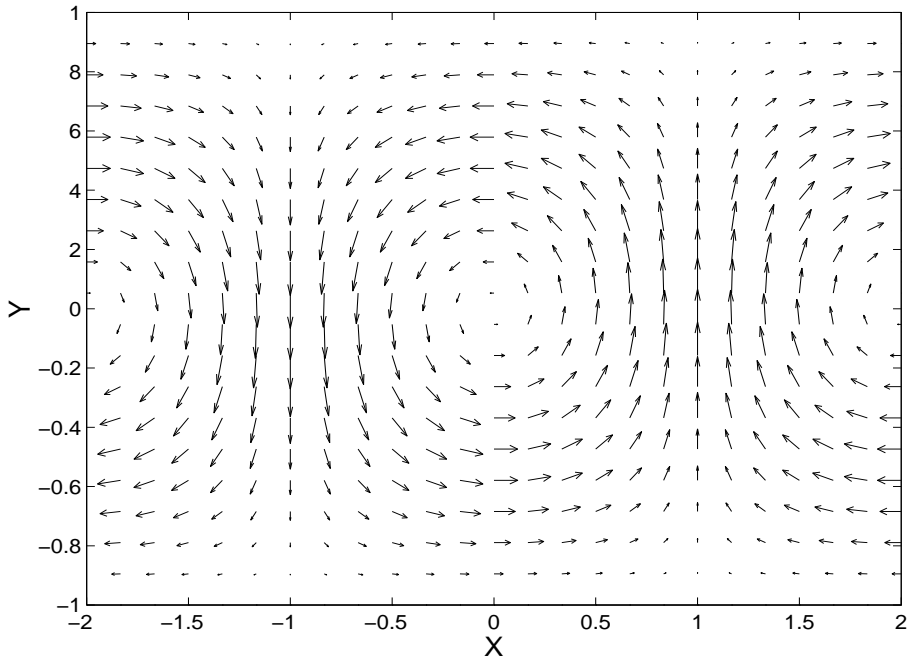}
  \caption{Velocity field: ${\bf v} = \nabla\psi{\times}\hat{{\bf
e}}_z$ using $\psi$ from (\ref{eq: sf}) with $k = \pi/2$, $\lambda = 2.6424$ and $A_{k\lambda} = .3499$}
  \label{fig:vfield}
  \end{center}
\end{figure}
\begin{figure}[ht]
  \begin{center}
  \includegraphics[width=\textwidth,height=3.5in]{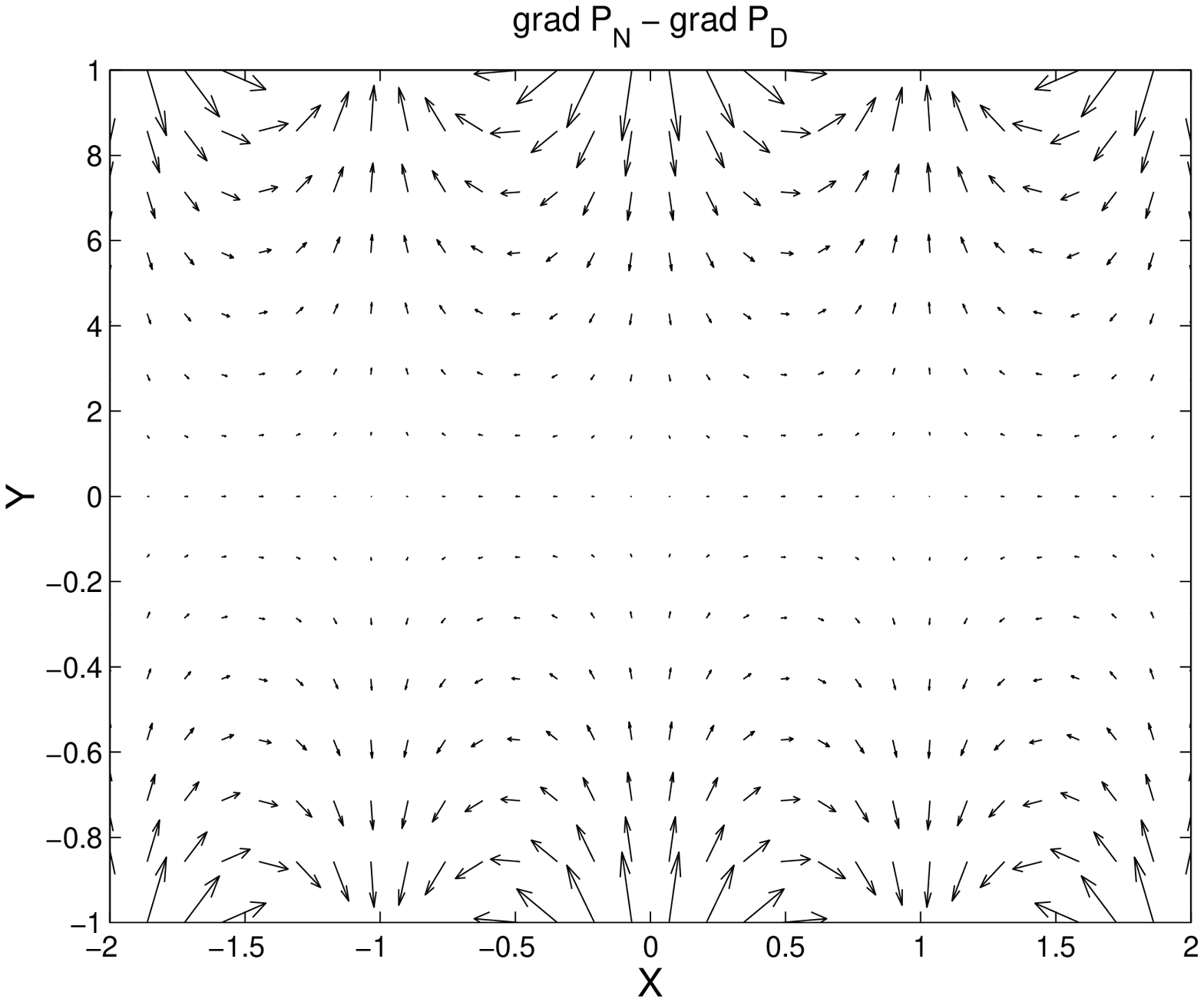}
  \caption{The difference between pressure gradients: $\nabla P_N-\nabla P_D$  with $\Rey=2293$}
  \label{fig:gradp}
  \end{center}
\end{figure}

We illustrate, in figure~\ref{fig:vfield}, an arrow plot of the velocity 
field given by choosing $k=\pi/2$, $\lambda=2.6424$ and $A_{k\lambda} = .3499$, in units of $a = 1$. The two pressures resulting from the satisfaction of the normal and tangential components of (\ref{eq: bcs}) can best be compared by 
comparing their respective values of $\nabla P$, since $P$ itself is 
indeterminate up to an additive constant in both cases. In 
figure~\ref{fig:gradp}, we display, as an arrow plot, the 
difference between the pressure gradients associated with the velocity field shown in figure~\ref{fig:vfield}. We have rewritten (\ref{eq: eqmo})-(\ref{eq: bcs}) in dimensionless units for this purpose, with the kinematic 
viscosity being replaced by the reciprocal of a Reynolds number, which may 
be defined as $\Rey=(\langle{\bf v}^2\rangle/(k^2+\lambda^2))^{1/2}/\nu$. Here, the angle brackets refer to the mean of ${\bf v}^2$ taken over the 2-D box, containing one period in the $x$ direction and from $y=-a$ to $y=a$. The value 
of \Rey\ used to construct figure~\ref{fig:gradp} is 
$\Rey\ =2293$, with the dimensionless version of $C_{k\lambda}=5000$ in (\ref{eq: sf}). The two pressures are similar but not identical.

\begin{figure}[ht]
  \begin{center}
  \includegraphics[width=\textwidth,height=3.5in]{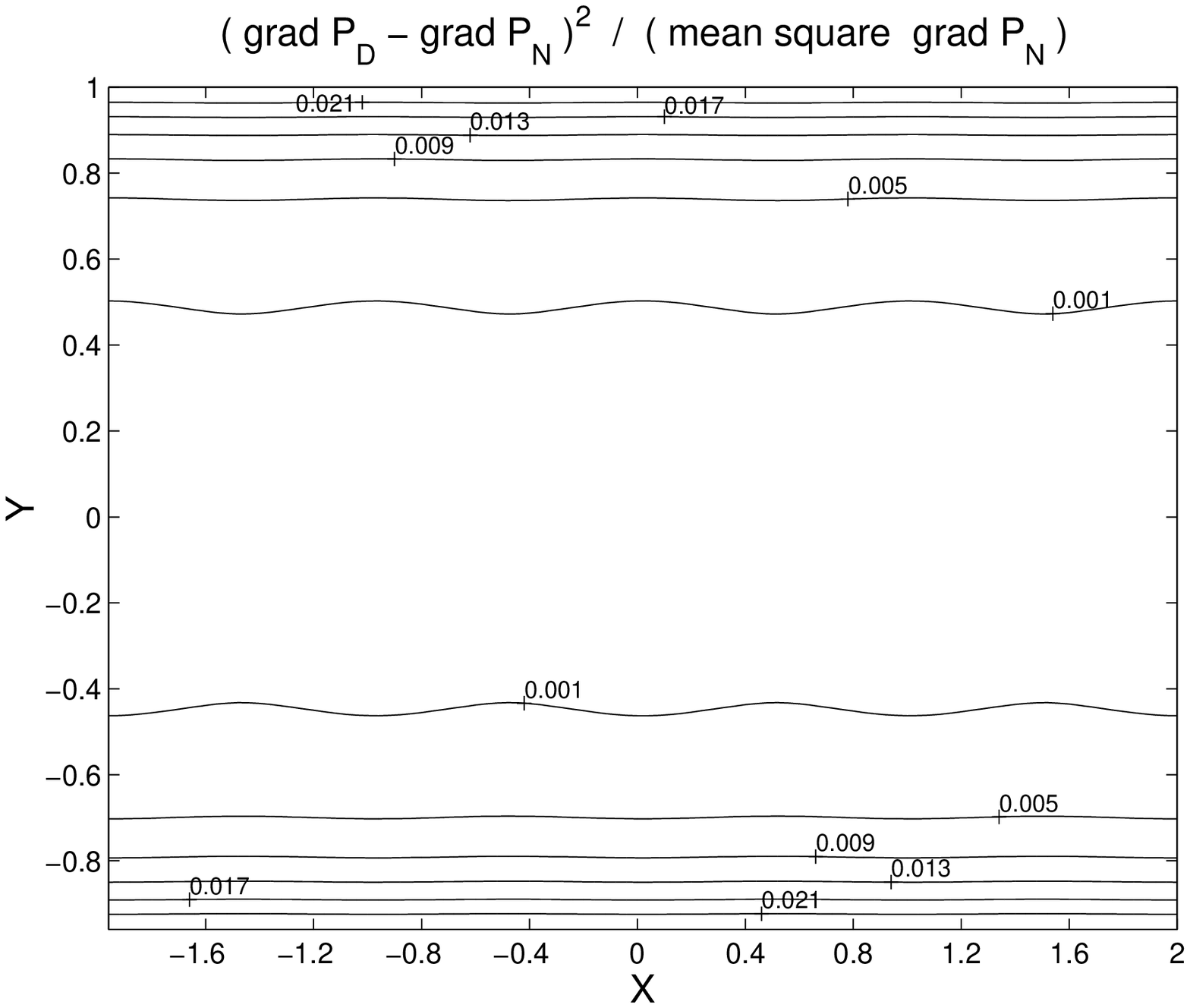}
  \caption{Normalized mean square pressure gradient difference: $(\nabla P_D-\nabla P_N)^2/\langle(\nabla P_N)^2\rangle$  with $\Rey=2293$. Note that the fractional difference between the two values of $\nabla P$ is significant only near the wall.}
  \label{fig:contour}
  \end{center} 
\end{figure}

In figure~\ref{fig:contour}, a fractional measure of the difference between the 
``Neumann pressure" $P_N$ and the ``Dirichlet pressure" $P_D$ is exhibited as a contour plot of the scalar ratio
\begin{equation}
\frac{(\nabla P_D-\nabla P_N)^2}{\langle(\nabla P_N)^2\rangle}
\label{eq: cplot}
\end{equation}
There is no absolute significance to the numerical value of this ratio. It initially increases with \Rey\, approaching a maximum of about $2\%$ near the wall for $\Rey \agt 10$. It is considered interesting however, that the fractional difference is nearly x-independent where it is largest.  That occurs formally because the algebra reveals it to be dominated by a term which varies as $\cosh{(4ky)} - \cos{(4kx)}$ in a region where $ky \agt 1$.

\begin{figure}[ht]
  \begin{center}
  \includegraphics[width=\textwidth,height=3.5in]{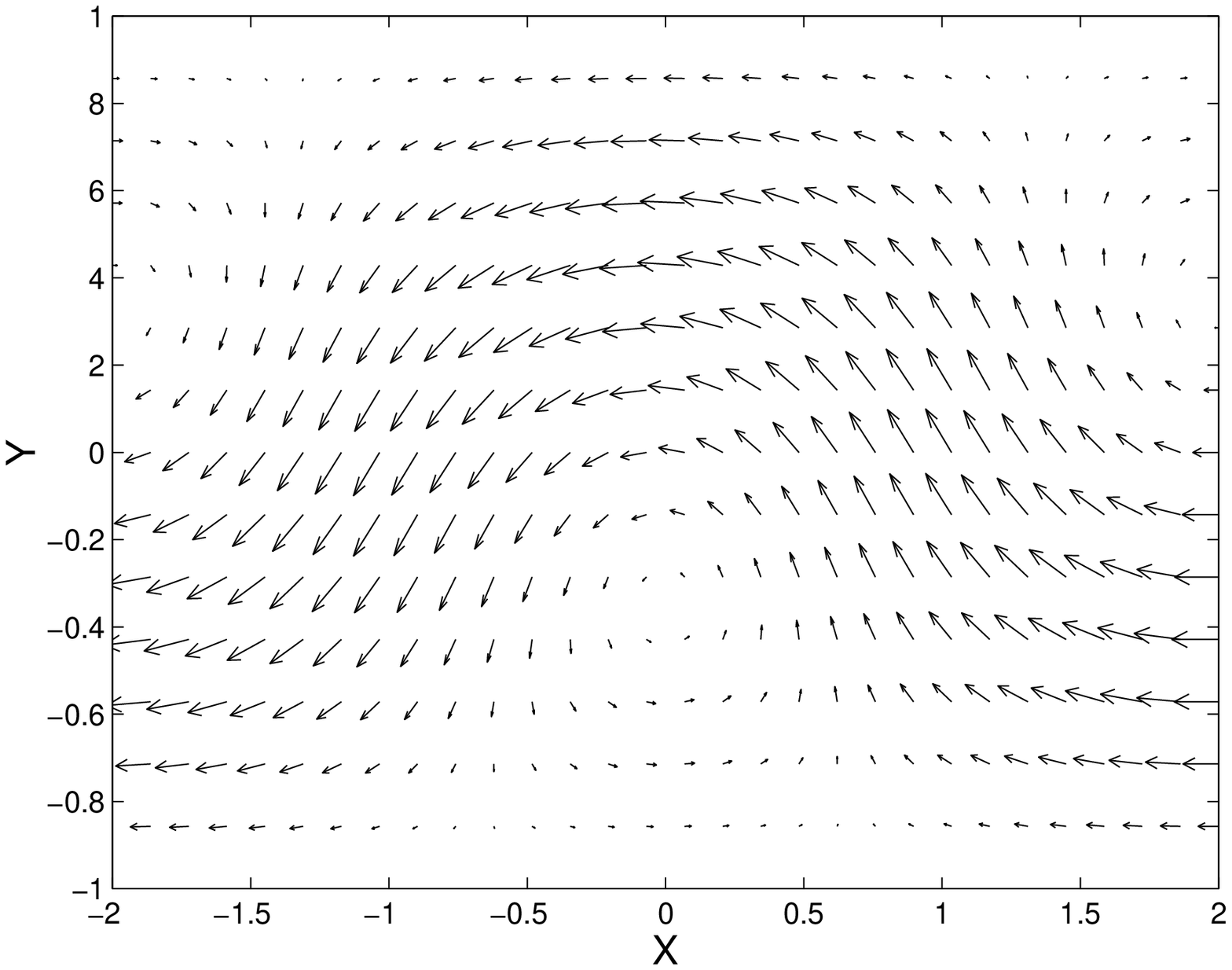}
  \caption{Velocity field shown in figure~\ref{fig:vfield} plus parabolic plane Poiseuille flow: ${\bf v} = \nabla\psi{\times}\hat{{\bf e}}_z+2(y^2-1)\hat{{\bf e}}_x$ with $C_{k\lambda}=1$}
  \label{fig:vstreet}
  \end{center}
\end{figure}

It is amusing but perhaps not significant to superpose the velocity field 
from (\ref{eq: sf}) with a parabolic plane Poiseuille flow of a larger 
amplitude. The resulting flow field is shown in figure~\ref{fig:vstreet}, 
and it bears a striking but perhaps not significant similarity to the flow 
patterns seen in two-dimensional plane Poiseuille flow \cite{Jones94} when 
linear stability thresholds are approached. The pressure gradient 
difference for this case will be fractionally smaller than in 
figure~\ref{fig:contour}, since pure parabolic plane Poiseuille flow is a 
rare case where the two pressures happen to agree, and it quantitatively 
dominates the pressures determined from equation (\ref{eq: pe}) in this example.

\section{Discussion and a possible modification}\label{sec:discussion}

An alternative to the no-slip condition is the ``Navier" boundary condition \cite{lamb32}: the slip velocity at the wall surface is taken to be proportional to the rate of shear at the wall. This may be expressed $\Delta V=L_s\dot{\gamma}$ where $\Delta V$ is the slip velocity of the fluid at the wall, $\dot{\gamma}$ is the rate of shear at the wall and $L_s$ is a constant with the dimensions length. Molecular dynamic simulations of Newtonian liquids under shear \cite{thompson90} have shown this to be the case under some circumstances. In fact recent work \cite{thompson97} has shown that, in cases where the shear rate is large, there is a nonlinear relationship between $L_s$ and $\dot{\gamma}$. 

We note that the velocity field shown in figure~\ref{fig:vfield} does not lead to one which obeys the Navier boundary condition, after an initial time step, where the fluid has been allowed to slip at the wall. If the velocity field determined by (\ref{eq: sf}) is advanced in time using (\ref{eq: eqmo}) with the ``Neumann pressure", the proportionality between the slip velocity and the rate of shear at the wall, after the initial time step, varies sinusoidally with $x$.

It is to be stressed that we are concerned here only with initial conditions, not with circumstances under which initial slip velocities might be coaxed dynamically into vanishing after some time.

It is difficult to see in what sense the velocity field obtained from  
(\ref{eq: sf}) might be an unacceptable one from the point of view of the 
Navier-Stokes or MHD descriptions. It seems to have all the properties that are thought to be relevant. The family of functions of the same x-periodicity in (\ref{eq: sf}) can be shown to be orthogonal, and is a candidate for a complete set, in which any ${\bf v}$ might be expanded, when supplemented by flux-bearing functions of y alone. The mathematical question of which if any 
velocity fields, which are both solenoidal and vanish at the wall, would 
lead to Neumann and Dirichlet pressures that were in agreement with each 
other, must remain open. Indeed, the question of whether there are any, 
without some degree of ``pre-processing," must remain open. This is an 
unsatisfactory situation for fluid mechanics and MHD, in our opinion, even if it is a not unfamiliar one. The search for alternatives seems mandatory.

One alternative that may be explored is one that seemed some time ago, in 
a rather different context [Shan \& Montgomery 1994a,b], to have worked well enough for 
MHD. Namely, we may think of replacing the requirement of the vanishing of 
the tangential velocity at a rigid wall with a wall friction term, added to 
the right hand side of (\ref{eq: eqmo}), of the form
\begin{equation}
-\frac{\bf v}{\tau({\bf x})}
\label{eq: dragterm}
\end{equation}
where the coefficient $1/\tau({\bf x})$ vanishes in the interior of the fluid and 
rises sharply to a large positive value near the wall. The region over 
which it is allowed to rise should be smaller than the characteristic 
thickness of any 
boundary layer that it might be intended to resolve, but seems otherwise 
not 
particularly restrictive. Such a term provides a mechanism for momentum 
loss to the wall and constrains the tangential velocity to small values, 
but does not 
force it to zero.  The Dirichlet boundary condition disappears in favor 
of a relation that permits the time evolution of the tangential components 
of ${\bf v}$, while demanding that $P$ be determined solely by the Neumann 
condition (the normal component of (\ref{eq: bcs}) only). In a previous 
MHD application [Shan \& Montgomery 1994a,b] dealing with rotating MHD fluids, the scheme 
seemed to perform acceptably well, but was not intensively tested or 
benchmarked sharply against any of the better understood Navier-Stokes 
flows. This comparison seems worthy of future attention.

The work of one of us (D.C.M.) was supported by hospitality in the Fluid Dynamics Laboratory at the Eindhoven University of Technology in the Netherlands. A preliminary account of this work was presented orally at a meeting of the American Physical Society \cite{Kress99}.

\end{document}